# Comments on the repulsive term of Van der Waals equation of state

Jorge Garcés

*Centro Atómico Bariloche, CNEA, 8400 Bariloche, Río Negro, Argentina*

The repulsive term of Van der Waals equation of state has been deduced and improved in this work using a probabilistic description of the configurational entropy. The aim is to find out whether its physical basis is suitable for the study of dense fluids like liquids and glasses. The extended Van der Waals model accurately describes a system composed of independent Hard Spheres below packing fraction $\eta < 0.3$, giving a pole at $\eta \sim 0.51$ in close agreement with Molecular Dynamic simulation results. The inclusion of holes in the configurational entropy description improves the model but only below $\eta < 0.2$, precluding its application to high density fluids and glasses. In addition, neither local nor global clustering of Hard Spheres can be described using this approach. A physical model other than Van der Waals equation of state is required to accurately describe the continuity of gaseous and liquid states with the same expression. In this work, a model for the configurational entropy of mixing is proposed which could reveal whether such continuity really exists.

## I. INTRODUCTION

The study of fluids began several centuries ago. The first efforts to describe the behavior of gases had an empirical nature since a theoretical foundation was unavailable at that time. One of the first successful attempts was due to Boyle in 1660 [1]. The first correction to explain the departures of real gases from Boyle's law was proposed by Bernoulli in 1738 [2], who suggested that the finite volume of molecules should be included in the model. This proposal was later resumed by Amagat [3], Hirn [4] and Dupré [5] in the nineteenth century. Hirn proposed the following expression,

$$(P + \theta)(V - v_0) = RT \qquad (1)$$

where $v_0$ is the sum of the volumes of the molecules. It was called co-volume by Dupré [6,7]. A significant advance was made by Andrews [8] who introduced new concepts such as the notion of continuity of the liquid and gaseous state and the critical phenomena in gases, the former being the central subject of the Van der Waals thesis [9]. These new ideas enabled this author to propose his well-known empirical equation of state (EOS) in 1873, which was considered a remarkable improvement on Boyle's law. The success of Van der Waals proposal was related to two approximations: i) the identification of the functional form of $\theta$, assumed to be $a/V^2$ and ii) the co-volume, also called excluded volume, was chosen to be four times the volumes of the molecules and not equal to $v_0$ as previously proposed. The Van der Waals equation of state (VdW EOS) has the following well-known expression,

$$(P + \frac{a}{V^2})(V - 4v_0) = RT \qquad (2)$$

or in P(V,T) variables as,

$$P = -\frac{a}{V^2} + \frac{RT}{(V - 4v_0)} \qquad (3)$$

VdW EOS described qualitatively for first time deviations from ideal gas, the existence of critical points and the gas-liquid transition. It has been widely applied in the study of gases and liquids, in spite of its approximate nature [10-12]. The empirical modifications used to adjust the experimental data were mainly carried out on the attractive first term in Eq. (3), while the repulsive second term remained unchanged [10-15]. In addition, none of the modified EOS properly describes the high-density region showing a divergence in unphysical limits, i.e. $\eta = 0.25$, far away from the actual value at Random Close Packing limit (RCP). Unfortunately, theoretical improvements to the VdW EOS are very difficult to achieve due to the empirical nature of the model. As a consequence, the technological research carried out in the first half of the 20$^{th}$ century was based on models and EOS with fundamentals that remain unexplained: e.g., in relation to the repulsive term of VdW EOS.

Why was the scientific research carried out using models with an unphysical basis? The answer is related to limitations in the theoretical methods available at that time, which were unable to describe the complex phenomenology of fluids for medium and high packing fractions, i.e. real gases, liquids and glasses. Consequently, simplified empirical models had to be used in order to advance in the understanding of the subtle, intriguing physics of fluids. The development of Statistical Mechanics models [16,17] and Molecular Dynamics simulations [18-21], introduced in the second half of the 20$^{th}$ century, enabled new physical phenomena to be found, stimulating a renaissance in scientific research on fluids and amorphous materials.

It is well known that limitations in theoretical scientific research can be overcome once new concepts and methodologies are introduced. The progress made in research on fluids over the centuries is one clear example. Another is the development of thermodynamic models of fluids, being the major obstacle the accurate calculation of the configurational entropy. Indeed, the traditional method using lattice gas model [22] is only suitable for applications to crystalline systems. An enormous amount of work has been done on the modeling of non-crystalline systems with the lattice gas model; e.g Flory [23], Huggins [24] or Gibss-DiMarzio [25] expressions, to cite only the most widely known models. However, deduction of an accurate and general expression for the configurational entropy of fluids that is valid for all densities has been very elusive. The lack of periodicity of atomic structures in fluids, and their changes under densification makes counting the number of configurations an almost insurmountable task.

However, an alternative formalism to compute configurational entropy has recently been introduced, providing a new theoretical framework for the development of accurate thermodynamic models [26,27]. Unlike the traditional method, which computes the number of configuration by counting the occupation in each node of the lattice, the new methodology is based on the identification of complexes, i.e. quasi-particles or any other suitable denomination, with small or null interaction energy between them and the calculation of their corresponding probabilities,

$$S = -k_B \sum_i n_i \ln p_i \tag{4}$$

where $n_i$ and $p_i$ are the number and probability of each independent complex $i$ in the mixture, respectively.

This formalism was successfully applied to a long-standing unsolved problem in Material Science: the development of an accurate, general expression for the configurational entropy of mixing of interstitial solid solutions [28]. However, this methodology has not yet been applied to the deduction of a similar expression for non-crystalline systems. The ultimate goal in such research is to find an adequate thermodynamic framework within which to study the glassy state with the same level of accuracy as gases and liquids.

The purpose of this work, as a first step in this research, is to use the formalism of Eq. (4) to identify the physical bases of VdW EOS repulsive term. In doing so, the model can be extended to high packing fractions to determine whether their fundamentals really allow the study of dense fluids, like liquids and glasses. The deduction of a universal expression for the configurational entropy of mixing of crystalline and non-crystalline systems is outlined here. A detailed calculation will be presented in a future study.

This work is organized as follows. Section II will be related to deduction of the extended Van der Waals model and its EOS. Results and comparisons with the well-studied non-interacting Hard Spheres (HS) system are presented in Section III. Limitations and improvements to the model are discussed in Section IV. Conclusions are presented in Section V.

## II. THE EXTENDED VDW EOS

The repulsive term of VdW EOS is difficult to improve due to the empirical deduction of the model. Nevertheless, some researchers have tried to rationalize the origin of this term. Of these, Rusanov [29-31] was able to improve previous EOS, such as those due to Planck [29,32] and Van der Waals [9], maintaining their simplicity by assuming dependence of the excluded volume with the packing fraction. However, to improve the Rusanov model it is necessary to know the dependence of clustering degree on fluid density, a theoretical problem which remains unsolved in spite of the efforts of many researchers. It will be assumed in this work that the excluded volume has an unknown dependence on density. It will be adjusted to MD simulation results for the HS system. In this way, the functional dependence of the excluded volume can be determined for all packing fractions.

### 1. The model

The assumptions required to deduce the repulsive term of vdW EOS from a free energy model based on Eq. (4) are: i) all $n$ atoms or molecules, considered as non-interacting HS entities with volume $v$, are independent, ii) there is no clustering of HS and iii) following Van der Waals's ideas, there is an excluded volume, called $\xi$ in this work, around each HS where occupancy of other HS is inhibited. Thus, the probability of finding any HS in an accessible volume $V$-$nv$-$\xi$ is computed by,

$$p = \frac{nv}{V - nv - \xi} \tag{5}$$

The volume $V$-$nv$-$\xi$ represents the volume available to the HS in which they have free movement. The total excluded volume $\xi(T,V)$ is an unknown function of the distribution of $n$ particles and holes in volume V. The corresponding entropy from Eq. (4) is,

$$S = -k_B n \ln\left(\frac{nv}{V - (nv + \xi)}\right) \tag{6}$$

The free energy for a non-interacting system is

$$F = -TS = nk_B T \ln\left(\frac{nv}{V - (nv + \xi)}\right) \tag{7}$$

The pressure is defined by,

$$P = -\left.\frac{\partial F}{\partial V}\right|_{T,n} = -\left(\frac{\partial(-TS)}{\partial V}\right) = T\frac{\partial S}{\partial V} \tag{8}$$

and is easily computed from Eqs. (6) and (8),

$$P = -nk_B T \left[\frac{1}{V - nv - \xi}\left(1 - \frac{\partial \xi}{\partial V}\right)\right] \tag{9}$$

Taking into account the following relation,

$$P = T\frac{\partial S}{\partial V} = T\frac{\partial S}{\partial \eta}\frac{\partial \eta}{\partial V} = -T\frac{\eta}{V}\frac{\partial S}{\partial \eta} \tag{10}$$

the compressibility factor Z can be deduced as a function of the packing fraction $\eta = nv/V$,

$$Z = \frac{PV}{nkT} = \frac{1}{1 - \eta(1 + \frac{\xi}{nv})}\left(1 + \eta^2 \frac{\partial(\xi/nv)}{\partial \eta}\right) \tag{11}$$

This equation shows for the first time the root causes of inaccuracies and limitations of most EOS available in the literature [11,12,33]. All of them omitted the hitherto unknown term $\partial(\xi/nv)/\partial\rho$, an important contribution for high density gases, liquids and glasses.

In this work, Eq. (11) will be called `extended VdW EOS`. This equation constitutes an improvement to the repulsive term of VdW EOS, as it explicitly includes the functional dependence of excluded volume $\xi/nv$ on the packing fraction. The theoretical methodology currently available does not provide clues as to how $\xi/nv(\eta)$ can be accurately deduced analytically for dense fluids. Consequently, a function with as few parameters as possible will be used to describe its functional dependence. A sigmoid function with three parameters is assumed for this purpose,

$$\frac{\xi}{nv} = \frac{A}{1+\exp(B(\rho-C))} \quad (12)$$

The MD results of Wu and Sadus will be used to adjust the parameters [20].

## III. RESULTS.

This section will study the behavior of Eq. (11) for different adjustments of parameters of excluded volume $\xi/nv$. MD simulation results available in the literature will be used to understand the physics described by each of the resulting EOS.

*A. One-parameter EOS: vdW EOS.*

The repulsive term of VdW EOS is recovered from Eq. (11) if a constant value $\xi/nv = 3$ is chosen for the full range of packing fractions,

$$Z = \frac{PV}{nkT} = \frac{1}{1-b\eta} = \frac{1}{1-4\eta} \quad (13)$$

The constant value $b = 4$ represents a coarse approximation to the description of dense fluids, as is well known. Eq. (13) has a pole at $\eta = 0.25$, far from the RCP limit, as shown in Fig. 1. Most of the empirically modified cubic EOS using Eq. (13) is applied beyond its validity limit requiring at least seven, and usually ten or more parameters, to describe real fluids across the range of packing fractions $0 < \eta <$ RCP [11,12,33].

*B. Two-parameter EOS: extended vdW EOS.*

It is assumed that vdW EOS accurately describes the limit of low packing fractions using the value $b = 4$. This value will be used as a constraint to fix one of the three parameters in Eq. (12). The other two will be determined by fitting Eq. (11) to the MD data of Wu and Sadus [20]. The values of the three parameters are A= 3.6787 (fixed by the dilute limit condition), B = 4.958 and C = 0.29975, from the best fit to MD data.

Fig. 1 compares the results of the extended VdW EOS with the MD results of Wu and Sadus. Also shown are Eq. (13) and the widely-used empirical EOS of Carnahan and Starling (CS) [34]. Fig. 2 shows the differences between Wu and Sadus MD data with Eq. (11), VdW EOS, and CS EOS. This figure shows that the repulsive term of VdW EOS gives a poor description even for low packing fractions. Although it is not the purpose of this work to compare current results to previous, well-established empirical EOS, it is seen that the adjustment provided by the extended vdW EOS is better than CS in the range $0 < \eta < 0.30$. The difference when compared with MD data is negligible.

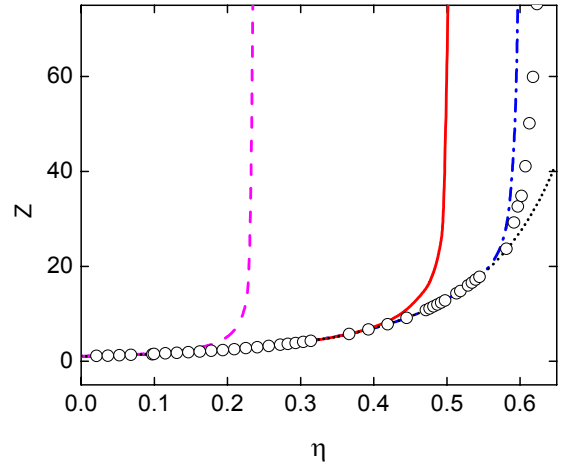

Figure 1. Compressibility factor Z vs. packing fraction η. Van der Waals EOS: magenta dash line. Extended VdW EOS with two parameters: red solid line. Extended VdW EOS with three parameters: blue dash-dot line. Carnahan and Starling EOS: dot line.

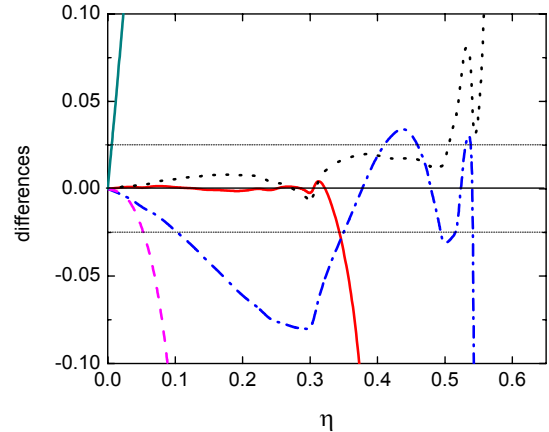

Figure 2. Difference between Wu and Sadus MD data for the compressibility factors and different EOS. Van der Waals EOS: magenta dash line. Extended VdW EOS with two parameters: red solid line. Extended VdW EOS with three parameters: blue dash-dot line. Carnahan and Starling EOS: dot line. Ideal gas law: green solid line.

Fig. 1 reveals a surprising outcome as the extended VdW EOS diverges at $\eta \approx 0.51$ and not at the RCP limit, as was the initial expectation in this research.

There is only one reference in the literature explaining the existence of a pole at $\eta \approx 0.51$ in the HS system. Blétry and Blétry [35] used two different algorithms in their MD simulations to control the location of the HS inside a reference sphere. One of them constructs the aggregate at random in the sphere, i.e. the HS are independent and there are no clustering effects, i.e. the same hypotheses used to deduce the extended VdW EOS. Those authors found that the random algorithm has an upper bound at $\eta \approx 0.51$. This limit has a remarkable agreement with the pole of the extended VdW EOS with two independent parameters. It is noticeable that a simple expression as Eq. (11) can reproduce the same limit as a sophisticated model such as MD using the same hypotheses. Although both models describe the same structural behavior of HS below $\eta < 0.51$, i.e. a random distribution of independent HS, the MD data of Wu and Sadus are accurately described by the extended VdW EOS only below $\eta \approx 0.3$.

*C. Three-parameter EOS.*

It can be argued that a better description for a wider range of packing fractions could be achieved using three parameters, i.e. discarding the dilute constraint $b = 4$. Figs. 1 and 2 show the results of fitting with three parameters. The parameters best fitting the MD data are: $A = 4.0131$, $B = 4.94405$ and $C = 0.27361$.

The fitting procedure highlights another unexpected result. Indeed, the EOS with three parameters adjusted well at medium packing fractions, showing a singularity at $\eta = 0.606$ and not at the RCP limit. This EOS gives a worse description of the MD data than CS in the dilute zones, and is comparable in the range $0.35 < \eta < 0.54$. MD data beyond $\eta = 0.545$ cannot be included in the EOS with any combination of the three parameters. Instability is developed if the parameters are forced to diverge at the RCP limit.

Unlike the pole $\eta \approx 0.51$, the pole at $\eta \approx 0.606$ is well documented in the literature. It can be understood by analyzing the experimental behavior of the HS system under densification. If a container is filled with HS, the spheres can be situated as compactly as possible or as detached as possible. These limits are called Random Close Packing (RCP) and Random Loose Packing (RLP) respectively [36]. Scott found experimentally that the RLP limit is located at $\eta_{RLP} = 0.608$, in remarkable agreement with the pole of the extended VdW EOS adjusted with three parameters.

MD results explain the circumstances under which this limit can be achieved:

*i)* Silbert [37] commented that the limit $\eta_{RLP} \approx 0.55$ can be reached only in the case of infinite friction between spheres. The value 0.608 can be achieved only for spheres without friction and under the effect of the gravitational field.

*ii)* Blétry and Blétry [35] remarked that packing fractions larger than 0.59 can not be reached by their algorithms which optimize the clusterized sphere positioning up to the second neighbor distance.

*iii)* Other authors using geometrical algorithms did not go beyond $\eta = 0.603$ [38] and $\eta = 0.6053$ [39]. This conclusion is also confirmed in Refs. [40] and [41].

Several MD simulations explain the structural behavior of the fluid near the RLP limit:

*iv)* Aste et al. [42] reported that beyond $\eta \approx 0.6$, densification can occur by collective readjustments of larger sets of spheres. MD static methods cannot describe this kind of behavior.

*v)* Anikeenko and Medveded [43] noted that the threshold value $\eta = 0.608$, corresponds to the density after which correlation of the disordered arrangement of spheres increases considerably. The relative fraction of the adjacent faces of tetrahedral-shape simplexes shows an inflection at $\eta = 0.60$. This feature is associated with changes in the topology of the tetrahedral clusters and reflects the increased correlations in the mutual arrangement of spheres in a packing.

*vi)* Tian et al [44] found that below the 0.608 limit the hard spheres are settled in what they called `local crowded structures`, characterized by local rearrangements that do not affect all HS. Beyond this limit there are `global crowded structures` characterized by rearrangements of all HS.

The remarkable description of the pole at $\eta_{RLP} \approx 0.60$ could lead to the supposition that the extended VdW EOS with three parameters correctly described the physics of the intermediate range of packing fractions and at the RLP limit. However, while the deduction of Eq. (11) is based on an independent HS behavior, MD results show that the intermediate zone is characterized by clustering of HS. It can be concluded that the extended VdW EOS with three parameters is not able to describe the structural feature of this intermediate packing fraction zone. A different approach able to include clustering of HS should be implemented for this zone and in the branch associated with the amorphous and glassy state.

*D. Excluded volume and percolation thresholds*

The behavior of the extended VdW EOS with two parameters below $\eta < 0.3$ and the pole at $\eta_{RLP} \approx 0.51$, in notable agreement with MD results, are important sources of confidence regarding the model provided by Eq. (11). Another source will arise from analysis of the geometrical structure of the excluded volume curves.

Fig. 3 shows the excluded volume curve from Eq. (11) with two parameters. The functions $\xi/nv$ and all their derivatives are continuous functions for all packing fractions. However, the figure clearly shows a change of curvature around $\eta \approx 0.30$. Analysis of the second derivative shows two extremes, one located at 0.035 and the other around $\eta \approx 0.30$, as shown in Fig. 4. The result implies that the fluid responds differently under densification in each of the three sections in which the full range of packing fractions is separated.

The values resemble those proposed in the literature for the percolation limits. Woodcock presented the first determination of values for the percolation threshold, although their existence had previously been recognized [45, 46]. This author reported $\eta = 0.0458$ and $\eta = 0.281$ for the excluded and available volume percolation thresholds, respectively, in close agreement with the results obtained using the extended VdW EOS with two parameters.

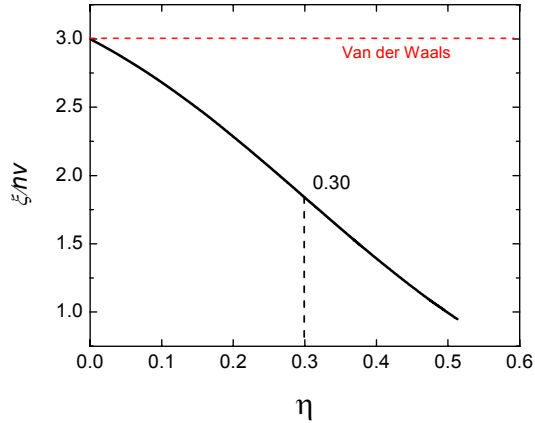

Figure 3. Excluded volume vs. packing fraction for the extended Van der Waals EOS with two independent parameters. The excluded volume shows a change of curvature at $\eta \approx 0.30$.

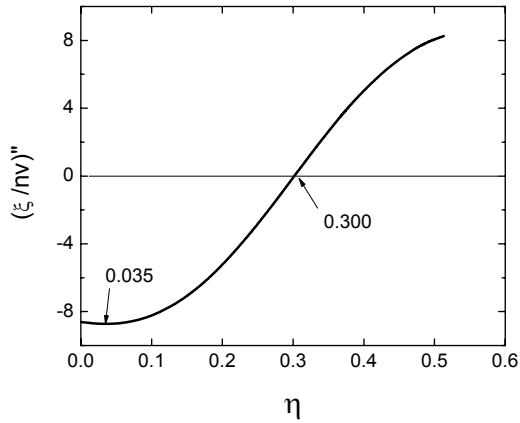

Figure 4. Extremes of the second derivative of the excluded volume vs. packing fractions $\eta$.

## IV. DISCUSSION.

At variance with the empirical proposal of VdW EOS, the extended vdW model is a theoretical deduction based on the configurational entropy calculation using the hypotheses leading to VdW repulsive term. Expectations at the beginning of this research were that the full range of packing fractions, $0 < \eta < RCP$, could be adjusted using that model. However, a remarkable adjustment could only be obtained below $\eta < 0.3$, presenting a pole at $\eta \approx 0.51$, in agreement with the MD results for independent HS of Blétry and Blétry [35]. It was also found that the foundation of the extended VdW EOS, i.e. random distribution of HS, precluded its application to states characterized by clustering of HS as in liquids and glasses. This is not the only limitation of the model. Indeed, this section will show that the configurational entropy leading to the repulsive term in VdW EOS is incomplete as it does not include the effects of holes distribution in its calculation. This is one of the main sources of inaccuracy in the description of dense fluids.

### A. The distribution of holes

As previously mentioned, the repulsive term of VdW EOS was used with no changes in almost all the cubic EOS available in the literature. One improvement made in this term led to the extended vdW model, which has the limitations commented on previously.

A second improvement to the repulsive term arises from the question: should the distribution of holes be included in the calculation of the configurational entropy? Eq. (6) is deduced without considering the effect of holes distribution. However, any realistic model valid for all packing fractions should include changes under densification of the different complexes in the mixture and the size and number of holes for all packing fractions. Therefore, the quantity to be computed is the configurational entropy of mixing a number $n$ of HS and a distribution of holes. That is a very difficult task to solve analytically. Some alternative approaches were implemented in the literature based on Voronoi partition of the space [47-49].

It is possible to include the contribution of holes using the formalism of Eq. (4) based only on the assumptions of the Van der Waals model, i.e. independent HS and excluded volume fulfilling the condition $b = 4$. For this purpose, the holes will be considered as independent quaiparticles, as in the deduction of an expression for the configurational entropy of mixing in interstitial solid solutions [28].

While the probability of finding any HS is the same as in the vdW model given by Eq. (5), the one for holes will be assumed simply as,

$$p^h = 1 - \frac{nv}{V-4nv} = 1 - \frac{\eta}{1-4\eta} = \frac{1-5\eta}{1-4\eta} \quad (14)$$

This assumption implies the subdivision of empty space in cells with volume $v$, as used in lattice gas models [22]. Consequently, the number of holes in this oversimplified model will be computed as the empty space divided by the HS volume,

$$n^h = \frac{V-4nv}{v} = \frac{V}{v}(1-4\eta) = n\left(\frac{1-4\eta}{\eta}\right) \quad (15)$$

The configurational entropy of mixing for a system of $n$ HS and $n^h$ holes is,

$$S = -k_B n \ln\left(\frac{nv}{V-4nv}\right) - \left(\frac{V-4nv}{v}\right)k_B \ln\left(1-\frac{nv}{V-4nv}\right) + C \quad (16)$$

Eq. (16) as a function of packing fraction $\eta$ is,

$$S = -nk_B\left[\ln\left(\frac{\eta}{1-4\eta}\right) + \frac{(1-4\eta)}{\eta}\ln\left(\frac{1-5\eta}{1-4\eta}\right)\right] + C \quad (17)$$

The constant C is needed to compare different models of entropy, such as those representatives of Eqs. (6) and (17), since entropy is undetermined to an arbitrary additive constant. Its value is determined in this work by adjusting both entropy models to the ideal gas limit.

Fig. 5 shows the comparison between the configurational entropy of ideal, VdW and extended VdW models and, the result of Eq. (17). The figure highlights the fact that the introduction of holes improves the VdW model above $\eta < 0.10$, i.e. the description is equivalent to the configurational entropy of the extended vdW EOS which exactly adjust the MD results. However, the range of applicability is reduced to $\eta < 0.18$ as a consequence of using the lattice gas model to compute the number of holes.

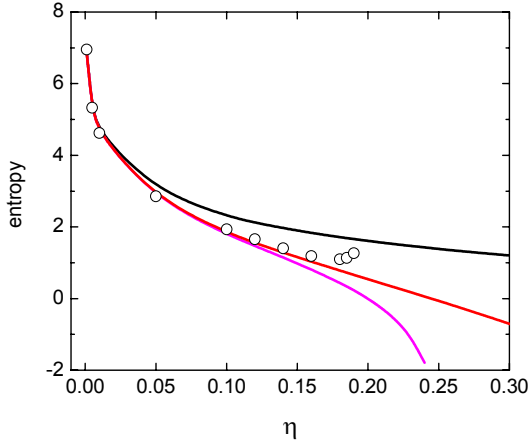

Figure 5. Configurational entropy vs. packing fraction $\eta$ of: a) Ideal: black line, b) extended VdW model: red line, c) VdW model from Eq. (6): magenta line. Eq. (17) including holes: empty circles. The constant C is adjusted to give the same entropy in the dilute limit. The value used in this work is $C = -0.95$

It is clear from the above results that: i) the VdW model can be improved by adding holes in the calculation of the configurational entropy, and ii) the excluded volume model, as used to deduce the VdW EOS with $b = 4$, will not accurately describe a dense fluid even if hole distribution is included in the formalism. A different approach is required for this purpose, as will be shown below.

*B. The formation of clusters*

The theoretical description of cluster formation in fluids is a troublesome task, mainly if their changes under densification and temperature are included in the formalism. There are many theoretical models which are useful to describe the local structure of the amorphous state [50-53]. However, it is still a challenge for theoretical research to encode all the information in a general expression for configurational entropy that can be applied to all non-crystalline states of matter. This goal cannot be achieved using the model developed in Section II because it is based on the concept of excluded volume and random distribution of HS; i.e. no clusters are taken into account in the fluid description. It is possible to add clustering effects in the model using the formalism of Eq. (4). However, VdW EOS cannot be derived from the final expression for the configurational entropy. This improvement leads to a complex expression, even more if holes are included in the calculation. This final expression is rather difficult to evaluate as it requires knowledge of the changes under densification of the excluded volume associated with each cluster. The calculation of both quantities constitutes an unsolved problem, as previously mentioned.

In this work, a different formalism for computing probabilities of clusters and holes for all packing fractions is implemented using the information available from MD simulations. For this purpose and as an example of the application of Eq. (4) to fluids, one feasible model for the configurational entropy of mixing of the HS system will be deduced using that information.

The system of interest is composed of $n$ non-interacting HS of volume $v$, capable of a geometrical cluster formation, and a variable holes distribution, both dependent on the total volume V. The configurational entropy of mixing for this system is given by,

$$S = -k_B \ln(P) = -k_B \ln\left(\prod_i p_i^{n_i}\right) = -k_B \left[\sum_i n_i^c \ln p_i^c + \sum_j n_j^h \ln p_j^h\right] \quad (18)$$

where $i$ applies to all clusters in the set. The probabilities will be computed as simply as possible, provided they contain all the relevant information about the system. In addition, the final expression for the configurational entropy will be required to include all previous expressions deduced using lattice gas models.

The probability of finding any cluster of the set is given by the packing fraction $\eta$, defined by,

$$\eta = \frac{\sum n_i v_i}{V} \quad (19)$$

The most general approach to compute the probability of finding a particular cluster of class $i$ is given by the conditional probability of finding the specific cluster $i$ once $\eta$ is known,

$$p_i^c(\rho,T) = \eta \, P_{cond}(cluster_i / \eta) \quad (20)$$

It is difficult to evaluate the conditional probability for a cluster $i$. It will be approximated in this work as independent from $\eta$ as,

$$P_{cond} \approx \frac{n_i^c(\eta,T)}{\sum n_j^c(\eta,T)} \quad (21)$$

Eq. (21) explicitly shows the functional dependence of the cluster distribution with packing fraction and temperature. Finally, the probability of a cluster $i$ will be computed as,

$$p_i^c = \eta \frac{n_i^c}{\sum n_j^c} \quad (22)$$

Similarly, the probability of holes is computed by,

$$p_i^h = (1-\eta)\frac{n_i^h(\eta,T)}{\sum n_j^h(\eta,T)} = (1-\eta)f_\gamma(k) \quad (23)$$

The probabilities of holes will be calculated as a Gamma distribution $f\gamma(k)$ of holes of sizes $v_i$ running from $v_{min}$ to $\infty$ [51-53],

$$f_\gamma(k) = \frac{k^k}{\gamma(k)}\left(\frac{v_i - v_{min}}{\overline{v} - v_{min}}\right)^{(k+1)} \exp\left(-k\frac{v_i - v_{min}}{\overline{v} - v_{min}}\right) \quad (24)$$

Finally, the configurational entropy of mixing is,

$$S = -k_B\left[n\ln\eta + \left(\sum_i n_i^h\right)\ln(1-\eta) + \sum_i n_i^c \ln\left(\frac{n_i^c}{\sum n_j^c}\right) + \sum_j n_j^h \ln f_j(k)\right] \quad (25)$$

It is important to mention here that Eq. (25) has a universal character, as can be applied to all off-lattice and lattice systems once the numbers and probabilities of each cluster under densification are given. In addition, it can be applied to multicomponent systems, the global composition of the systems constituting a constraint on the composition of all clusters in the mixture.

The first term of Eq. (25) is the main contribution in the diluted zone leading to the ideal gas law $PV = nRT$ or $Z = 1$.

Previous expressions for the configurational entropy of mixing based on the lattice gas model are included in the third term, which is the most important in a regular lattice as $\eta$ and $f_\gamma(k)$ have null effect in this system. To cite only the most relevant expressions in each field, this term includes expressions for polymers [23], lattice-glass models [25], interstitial solid solutions [27,28] and order-disorder in alloys using CVM method [54].

At this point some comments should be cited regarding the inclusion of clusters in the description of fluids using Eq. (25).

*i) What should the definition of clusters for the model of Eq. (25) be?*

Any combination of HS can be selected to define a cluster, but the criterion should be the same for every packing fraction and temperature. There are several possible choices for the local arrangement of HS in the literature, all of which recognize the existence of competition between tetrahedral and icosahedral symmetries. To cite only the most relevant to be implemented using the current formalism: i) the efficient packing cluster model of Miracle [52,53], ii) the topological cluster classification method of Royall and Williams [51], iii) cluster-plus-glue-atom model of Dong et al [55], iv) the description of Tian et al [44]. The last authors select a pair of HS and construct all the clusters around this figure according to the packing fraction. They provide all the information required to apply Eq. (25). This work is currently under development.

*ii) The continuity of gaseous and liquid states.*

Authors like Andrews, Van der Waals and others proposed the existence of a continuous EOS connecting gaseous and liquid states. VdW EOS led the researchers to believe that this was the model they required to explain the continuity and the existence of critical points. Indeed, Science, as a human activity, is influenced by the culture and scientific theories of the times. Therefore, a new paradigm was built based on those earlier ideas which were fruitful in guiding and inspiring a lot of research which helped to organize our current understanding of fluids.

The VdW model does not give a realistic description of real fluids. The results of this work show that the VdW and extended VdW EOS describe a hypothetical fluid composed of independent HS. It is known through MD results that gases and liquid phases have different local structures or clusters characterized by different distribution of HS [44]. Hence, the description of fluid through an EOS describing independent HS has no physical foundations and consequently, an accurate theoretical description of the gas-liquid continuity cannot be achieved using VdW EOS.

However, the idea of the continuity of liquid and gaseous states can be resumed based on Eq. (25) which takes into account the changes under densification and temperature of the different complexes in the mixture. The equation is simple but includes the most relevant characteristics of a fluid.

*iii) The vdW EOS critical points.*

It is known that in the vicinity of a critical point there are density fluctuations related to the formation of clusters of different sizes. Consequently, the physical meaning of a critical point arising from an EOS derived from an entropic term able only to describe independent complexes in the fluid has yet to be understood.

The existence of critical points is currently criticized by Woodcock [56,57]. Indeed, at variance with other researchers, based on his MD simulations results, this author proposed that there are no critical points. Instead, there is a two-phase field limited by the locus of percolation transitions. An accurate theoretical model is required to settle the difference between the critical point and the supercritical mesophase proposed by Woodcock. Eq. (25) provides a theoretical framework that could help to solve this controversy.

**V. CONCLUSIONS.**

The repulsive term of VdW EOS is improved in this work. The physical description of an independent HS system made by the extended VdW model is equivalent to those obtained from random geometrical algorithms using MD simulations [35]: neither local nor global clustering can be described using this approach.

The main conclusion of this work is that any model computing configurational entropy based on the assumptions leading to VdW or the extended VdW EOS cannot be used to describe dense systems like liquids or glasses. The reasons for this are: i) the clustering of HS is not considered and ii) the distribution of holes is not included in the calculation of the configurational entropy. Both effects are important in any accurate description of amorphous and glassy states. It is shown that the model leading to VdW EOS can be improved by adding the distribution of holes. However, the new model is not able to describe a dense fluid because the range of applicability is reduced to $\eta < 0.2$, if the assumption *b = 4* of VdW EOS is used.

The extended VdW model can be fitted accurately to MD data. This EOS is valid in the range $0 < \eta < 0.30$ having a pole at $\eta = 0.51$, in agreement with MD results of Blétry and Blétry [35]. The excluded volume vs. packing fraction clearly shows a change of curvature around $\eta \approx 0.30$, implying a change in the fluid response under densification. In addition, its second derivative shows two extremes located at 0.035 and 0.30. These values resemble those proposed by Woodcock for the percolation threshold [45]. This author reported $\eta = .0458$ and $\eta = 0.281$ for the excluded and available volume percolation thresholds, respectively.

Andrews, Amagat and Van der Waals conjecture related to the existence of a continuous EOS connecting gaseous and liquid states can not be attained using the hypotheses of the VdW EOS model: i.e., excluded volume and independent HS. However, the proposal cannot be disregarded based on this oversimplified model. A universal, general expression for the configurational entropy could be a better starting point for achievement of this goal. However, the main condition is to include in the formulation the ability to describe the change in local and global cluster populations vs. packing fraction and temperature. This approach, outlined in Section IV, is currently under development.

## REFERENCES

[1] R. Boyle. New Experiments Physico-Mechanical. Touching the spring of the air. 1660. Oxford: H. Hall,
[2] See Introduction by J.S. Rowlinson for a detailed historical reviews in: On the continuity of Gaseous and liquid states. Edited by J.S. Rowlinson. 1988 Dover Publications, Inc. Mineola, New York.
[3] E. H. Amagat. Sur la Compressibilitè de Gaz, Compt. Rendus, 71, 67, (1870)
[4] G.A. Hirn. Exposition Analytique et Expérimentale de la Théorie Méchanique de la Chaleur (CHez Mallet-Bachelier, Paris) 1862.
[5] A.L.V. Dupré. Théorie Méchanique de la Chaleur. (Gauthier-Villars, Paris) 1869.
[6] Jaime Wisniak. Educación Química 17, 86-96 (2006).
[8] T. Andrews. On the continuity of Gaseous and liquid states of matter. Phil. Trans. 159, 575-590 (1869).
[9] J.D. Van der Waals. Over the continuiteit van den Gas-en Vloeistoftoestand. Thesis. Univ. of Leiden. 1873; On the continuity of Gaseous and liquid states. Edited by J.S. Rowlinson. 1988 Dover Publications, Inc. Mineola, New York. Studies in Statistical Mechanics, vol. 14, 1988.
[10] J.O Valderrama. The state of th e cubic Euation od State. *Ind. Eng. Chem. Res*. **2003**, *42*, 1603-1618.
[11] A. Mulero, C.A. Galán, M.I. Parra and F. Cuadros. Chapter 3, pag. 37-109, in Theory and simulation of hard-sphere fluids and related systems. Lect. Notes Phys. 753 (Springer, Berlin Heidelberg 2008).
[12] S.M. Wallas. Phase Equilibria in Chemical Engineering. Butterworth Publishers, Boston. (1985)
[13] O. Redlich and J.N.S. Kwong, On the thermodynamics of solutions. Fugacities of Gseous solutions. Chem. Rev. 1949, 44, 233.
[14] G. Soave. Equilibrium Constants from modified Redlich-Kwong Equation of State. Chem. Eng. Sci. 1972, 27, 1197.
[15] ´D.Y. Peng and D.B. Robinson. A new two-constant equation of state. Ind. Eng. Chem. Fundam. 1976, 15, 59.
[16] J.K. Percus and G.J. Yevick. Analysis of classical statistic mechanics by means of collective coordinates. Phys. Rev. 110, 1-13 (1958).
[17] M.S. Wertheim. Analytical solutions of the Percus-Yevick Equation. Journal of Mathematical Physics 5, (1964) 643-651.
[18] B.J. Alder and T.E. Wainwright. Phase transition for a hard sphere system. J.Che. Phys. 31, 459 (1957).
[19] W.G. Hoover and F.H. Ree. Melting transition and communal entropy for hard spheres. J.Che.Phys. 49, 3609 (1968).
[20] Guang-Wen Wu and R.J. Sadus. Hard Sphere compressibility factors for equations of state development. Aiche Journal 51 (2005) 309-313.
[21] M.N. Bannerman, L. Lue and L.V. Woodcock. Thermodynamic pressures for hard spheres and close-virial equation-of-state. J. Chem. Phys. 132, 084507 (2010).
[22] J.S. Rowlinson. The rise and fall of lattice theories of the liquid state. Molecular Physics 113, 1-10 (2015).
[23] Flory, P. Thermodynamics of High Polymer Solutions. *J. Chem. Phys.* **1942**, *10*, 51-61.
[24].Huggins, M. Thermodynamics properties of solutions of long-chain compounds. *Ann. N.Y. Acad. Sci.* **1942**, *43*, 1-32.
[25] J.L. Gibbs and E. A. DiMarzio. J.Che. Phys. 28 (1958), 373-383.
[26] J.E. Garces. The configurational entropy of interstitials solid solutions. Appl. Phys. Lett. 96, 161904 (2010).
[27] J.E. Garces A probabilistic description of the configurational entropy of mixing. Entropy 16(5) (2014) 2850-2868.
[28] J.E. Garces. Short-range order of H in the Nb-H solid solution. Int. J. of Hydrogen Energy 39 (2014) 8852-8860.
[29] Rusanov, A. I. Equation of state theory based on excluded volume. *J. Chem. Phys.* **2003**, *118*, 10157-10163.
[30] Rusanov, A. I. Generalized equation of state and exclusion factor for multicomponent systems. *J. Chem. Phys.* **2003**, *119*, 10268-10273.
[31] Rusanov, A. I. Theory of excluded volume equation of state; Higher approximations and new generation of

equations of state for entire density range. *J. Chem. Phys.* **2004**, *121*, 1873-1877.

[32] M. Planck. Ueber die kanonische Zustandsgleichung einatomiger Gase. Sitzungsberg. *K. Preuss Akad. Wiss.* 32 (1908), 633-649.

[33] Thesis J.M. Garcia Palanco. Universidad Complutense de Madrid, 2013.

[34] N.F. Carnahan and K.E. Starling. Equation of state for nonattracting rigid spheres. J. Chem. Phys. 51, 635-636 (1969).

[35] M. Blétry and J. Blétry. Fluctuations, structure factor and polytetrahedra in random packing of sticky hard spheres. Journal of non-crystalline solids 411, 85-100 (2015).

[36] Scott, G. D., 1960, Nature, Lond., 188, 908-9; Brit. J. Appl. Phys. (J.Phys.D) ser2 vol 2, 863,(1969).

[37] L.E. Silbert. Soft Matter 6, 2918-2924 (2010).

[38] L.T.To and Z.H. Stachurski. Random close packing of spheres in a round cell. J. non.crystalline solids 333, 161-171(2004).

[39] R. Jullien and P. Meakin, Random sequential adsorption with restructuring in two dimensions. J. Phys.A: Math. And General 25, L189-L194 (1992).

[40] D. Aristoff and C. Radin. Random loose packing in Granular Matter. J. Stat. Phys. 135, 1–23 (2009).

[41] Jerkins Schröter, M., Swinney, H.L., Senden, T.J., Saadatfar, M., Aste, T.: Onset of mechanical stability in random packings of frictional particles. Phys. Rev. Lett. 101, 018301 (2008).

[42] T. Aste, M. Saadatfar, A. Sakellariou and T. Senden. Investigating the geometrical structure of disordered sphere packings, Physica A 339 (2004) 16.

[43] A. Anikeenko and N.N. Medveded. Polytetrahedral nature of the dense disordered packing of hard spheres. Phys. Rev. Lett. 98, 235504 (2007).

[44] Z.A. Tian, K.J. Dong and A.B. Yu. Structural evolution in the packing of uniform spheres. Phys. Rev. E 89, 032303 (2014).

[45] L.V. Woodcock. Percolation transitions in the hard-sphere fluid. Aiche Journal 58, 1610-1618 (2012)

[46] K.W. Kratky. Is the percolation transition of hard spheres a thermodynamic phase transition. J. Stat. Phys. 52, 1413-1421 (1988).

[47] Aste, T; Di Matteo, T. Emergence of Gamma distributions in granular materials and packing models. *Phys. Rev. E*. **2008**, *77*, 021309.

[48] 13. Aste. T.; Di Matteo, T. Structural transitions in granular packs: Statistical mechanics and statistical geometricy investigations. *European Phys. Journal B*. **2008**, *64*, 511-517.

[49] Kumar, V.S.; Kumaran, V. Voronoi cell volume distribution and configurational entropy of hard-spheres. *J. Chem. Phys*. **2005**, *123*, 114501.

[50] P. Ronceray and P. Harrowell. Favoured local structures in liquids and solids: a 3D lattice model. Soft Matter 11, 3322-3321 (2015).

[51] C.P. Royall and S. R. Williams. The role of local structure in dynamical arrest. Physics Reports 560, 1-75 (2015).

[52] D. Miracle. A structural model for metallic glasses. *Nat. Mater*. **2004**, *3*, 697-702.

[53] D. Miracle. D. The efficient cluster packing model – An atomic structural model for metallic glasses. *Acta Materialia* **2006**, *54*, 4317-4336.

[54] R. Kikuchi. A Theory of Cooperative Phenomena. *Phys. Rev*. **1951**, *81*, 988-1003.

[55] C. Dong, Q. Wang, J.B. Qiang, Y.M. Wang, N. Jiang, G. Han, Y.H. Li, Y. Wu and J.H. Xia. From Clusters to phase diagrams: Composition rules of quasicrystals and bulk metallic glasses. J. Phys. D: Appl. Phys. 2007, *40*, R273-R291.

[56] L.V. Woodcock. Nonexistence of a liquid-gas critical point. Science April 14, 1-12 (2012).

[57] J.L. Finney and L.V. Woodcock. Renaissance of Bernal`s random close packing and hypercritical line in the theory of liquids. J. of Physics: Cond. Matter 26, 463102 (2014)